\documentstyle[graphicx, amsmath, latin1]{mn}

\title{Funnel wall jets and the nature of the soft X-ray excess}

\author[N.J.\,Schurch \& C.\,Done]
{N.J.\,Schurch$^{*}$ \& C.\,Done\\ 
Department of Physics, Durham University, South Road, Durham, DH1 3LE, UK \\
$^{*}$ nicholas.schurch@durham.ac.uk}

\date{}

\def\H0{{\rm ~km~s^{-1}~Mpc^{-1}}}

\def\etal{et al.~\/}
\def\eg{{\it e.g.~\/}}

\def\la{\mathrel{\hbox{\rlap{\hbox{\lower4pt\hbox{$\sim$}}}{\raise2pt\hbox{$<$}}}}}
\def\ga{\mathrel{\hbox{\rlap{\hbox{\lower4pt\hbox{$\sim$}}}{\raise2pt\hbox{$>$}}}}}
\def\ls{\mathrel{\hbox{\rlap{\hbox{\lower4pt\hbox{$\sim$}}}\hbox{$<$}}}}
\def\gs{\mathrel{\hbox{\rlap{\hbox{\lower4pt\hbox{$\sim$}}}\hbox{$>$}}}}
\def\d25{D$_{\rm 25}$}

\def\.25{0.25 keV\thinspace}

\def\xst{{\small XSTAR}}

\def\xsc{{\small XSCORT}}

\begin{document}

\maketitle 

\begin{abstract}

The smooth soft X-ray excess seen in many type-1 AGN can be well described by models of absorption in partially ionized material with a large velocity dispersion, often physically interpreted as a radiatively driven accretion disk wind. However, the state-of-the-art \xsc ~code, which calculates the photoionized radiative transfer through a differentially outflowing absorber, shows that terminal velocities of order $\sim$0.9c are required in order to reproduce the soft X-ray excess. Such a high outflow velocity rules out UV-line driving, continuum radiation driving, and thermal driving as mechanisms for producing the wind. Entrainment of material by the magnetically driven jet is the only plausible origin of such a high velocity flow, but numerical simulations of jets and associated outflows do not currently show sufficient material at high enough velocities to reproduce the soft X-ray excess. If the soft excess is produced by absorption then it seems more likely that the material is clumpy and/or only partially covers the source rather than forming a continuous outflow.

\end{abstract}

\begin{keywords}
quasars: general, radiative transfer, accretion, accretion discs, X-rays: galaxies, galaxies: active, galaxies: Seyfert

\end{keywords}

\section{The mysterious soft X-ray excess}
\label{1}

The nature of the soft X-ray excess, observed in many type-1 AGN, remains a mystery, despite intensive study in recent years (\eg Porquet \etal 2004, Gierli\'nski \& Done 2004 - hereafter GD04, Crummy \etal 2006, Schurch \& Done 2006 - hereafter SD06). This smooth component shows no obvious features and can be modelled with a thermal continuum. However, this modelling reveals a characteristic `temperature' for the soft X-ray excess that remains remarkably constant across a range of type-1 AGN, despite a large spread in the intrinsic properties of the central engine (\eg Czerny \etal 2003, GD04). The lack of any clear dependence on the underlying source properties is very difficult to explain with either accretion disk (\eg Shakura \& Syunyaev 1973, Mineshige \etal 2000) or Comptonization models (\eg Porquet \etal 2004, GD04), particularly given the large dispersion in black hole masses for observed type-1 AGN (GD04). As such, it is unlikely that the soft X-ray excess is a true continuum component.

Atomic transitions in partially ionized material provide a simple mechanism for reproducing such a fixed temperature feature and, in particular, the strong O{\small VII/VIII} absorption between $\sim$0.7-3 keV provides a natural origin for the soft X-ray excess. Any atomic origin for this feature must, however, incorporate either considerable velocity smearing in order to blur the characteristically sharp individual features into a smooth continuum-like component. Simple models based on reflection and absorption from such material have both been successful in reproducing the characteristic shape and spectral variability of the soft X-ray excess and, as a result, have both been proposed as potential origins for this feature (\eg Fabian \etal 2002, 2005, Crummy \etal 2006, GD04, SD06, Chevallier et al. 2006, Sobolewska \& Done 2007, Middleton \etal 2007). However each of these scenarios encounters significant problems when examined in detail. In the case of reflection from the accretion disk, models of X-ray irradiated material in pressure balance show that the ionization instability limits the column density of material that can be present in the partially ionized state (Chevallier \etal 2006), and that this column is insufficient to dominate the entire photosphere of the accretion disc (Done \& Nayakshin 2007). In the case of absorption associated with an accelerating accretion disk wind, self consistent models that incorporate the velocity field into the photoionization calculation, show that a terminal velocity of $\sim$0.3c (the maximum expected for a UV-line driven disk wind in AGN - Proga \& Kallman 2004) produces insufficient velocity dispersion to remove the characteristically sharp atomic features associated with absorption in partially ionized material (Schurch \& Done 2007 - hereafter SD07).

This effectively rules out all current plausible models for the soft excess! Nonetheless, absorption in a high velocity outflow may still hold the key to understanding this mysterious feature. While thermal and radiation (both continuum and line) driving of an accretion disk wind are unable to produce a wind with sufficient velocity dispersion, there is also the potential for magnetic driving to produce a considerably faster outflow associated with the jet.  Studies of the magnetic fields associated with accretion flows have only recently become feasible (\eg Hawley \& Krolik 2006, Kato 2007), though as yet they cannot give a robust guide to the outflow properties. Here we simply assume that magnetic driving can produce higher velocity outflows, and examine the effect of these higher velocity outflows on an intrinsic AGN X-ray spectrum. We show that absorption by material in such an outflow can indeed reproduce the shape of the soft X-ray excess, but that the required outflow velocities are extreme, even for an association with the jet.

\section{\xsc ~spectra of highly relativistic winds}
\label{2}

The \xsc ~code (SD07) is a numerical code that links a series of radiative transfer calculations, incorporating the effects of a global velocity field in a self-consistent manner (including special relativistic effects), to produce model UV/X-ray spectra of AGN observed through the outflowing material. The \xsc ~spectra presented in SD07 take a power-law continuum as the initial ionizing spectrum and explore the absorption relative to a standard set of input parameters based, in part, on the set of `average' AGN wind properties from Middleton \etal (2007) (column of 3$\times$10$^{23}$ cm$^{-2}$, ionization parameter $\xi=L/nr^2=10^3$ where $L$ is the luminosity, $n$ is the number density and $r$ is the distance from the ionizing source) and on the hydrodynamic simulations of accretion disk winds from Proga (2000) and Proga \etal (2004) which suggest a maximum terminal velocity of $\sim$0.3c. The resulting spectra do not reproduce the observed smoothness soft X-ray excess, primarily because the resulting velocity dispersion in these simulations is insufficient to smear the sharp atomic features into a smooth continuum-like component.  Here we calculate a series of \xsc ~models with considerably higher terminal velocities in order to see whether these can indeed match the observed characteristics of the soft excess.

The increase in velocity dramatically enhances the impact of absorption lines on the spectrum. For any given velocity width (from either ordered acceleration or turbulence), the equivalent width of a line saturates with increasing column as the line becomes optically thick, limiting its value. For a given saturated line, increasing the velocity increases the intrinsic line width, and so decreases the optical depth in the line core. However, for lines that are very optically thick, the core of the line remains black irrespective of whether the optical depth is 10 or 10$^{4}$. Increasing the velocity width thus has the effect of increasing the intrinsic width of the saturated line core and results in an increase of the equivalent width of the line. However, once the velocities are large enough that the line is no longer optically thick at its core, the decreasing optical depth of the line counteracts its increasing line width, resulting in a constant equivalent width with increasing velocity.

The high column used as standard in SD07 produces an Fe L edge that extends the absorption out to $\sim$3 keV as required by the data, but also results in the O{\small VII/VIII} being saturated. The associated broad absorption trough at $\sim$0.7 keV is black, requiring in-filling from the associated emission in order to match the observed depth (but not smoothness) of the soft excess. For such high columns, simply increasing the terminal velocity of the outflow would increase the equivalent width of the O{\small VII/VIII} absorption line, making it harder to match the observed soft excess profile. This forces a reduction in the column density of the absorbing material for the high velocity winds presented here, which also has the effect of reducing the contribution of the emission spectrum (see \eg SD06), and the Fe L edges (which scale simply with N$_{H}$, with little dependence on velocity). The resulting spectral shape is governed, to first order, by the O{\small VII/VIII} absorption alone. We choose a column of 6$\times$10$^{22}$ cm$^{-2}$ so that the depth of the resulting broad absorption trough matches that of the simplistic average AGN wind models from Middleton \etal (2007), which are known to provide a good fit to a range of soft excesses in X-ray AGN spectra.

The large velocity also enhances the luminosity and density differences between the rest frame of the source and that of the outflow. Imposing a constant density on the material in the rest frame of the source then results in a strongly decreasing ionization parameter in the rest frame of the outflow even if the radius does not change significantly. As discussed in SD07, we instead impose a gas density structure on the wind material that directly counteracts these relativistic changes, so that (to first order) there is a constant ionization state throughout the material. Any remaining variation in the ionization parameter is small, due only the decrease in luminosity due to absorption by the wind material and the increase in distance of each shell from the ionizing source.

The highly relativistic terminal wind velocities simulated here introduce some additional numerical complexity to the \xsc ~calculations. For a typical mildly relativistic wind, the radiative transfer calculations are processed on a 100 step grid and, for a linear velocity law, the velocity shear across each shell produces individual features with widths of approximately the same order as the energy resolution of the internal \xsc~ grid (we note that this does not refer to the energy resolution of the internal \xst ~energy grid that defines the resolution of the spectrum). Thus, for a given individual feature, individual steps in the velocity grid are not resolved and smooth velocity smeared features are produced. For the considerably faster, more rapidly accelerating, winds presented here, however, such a course grid produces substantial sinusoidal wiggles in the resulting spectra, particularly around sharp features. We mitigate this effect as much as possible by increasing the number of steps in the \xsc ~chain however an important consequence of this is a corresponding increase in the errors introduced by the interpolation routines used to shift the component spectra into, and out of, the rest frame of each shell (SD07). These errors are particularly important when the width of narrow features in each \xst ~spectrum are comparable to the size of the energy grid. A large number of interpolations then act to fill in the absorption features and reduce the strength of emission features. In extreme cases, these errors can be very significant (factors of $\sim$3-4), however we minimise their impact on here by ensuring that the turbulent velocity used to calculate the spectrum from each sub-shell is always large enough that none of the spectral features are narrow in comparison with the energy grid.

\begin{figure}
\centering
\begin{minipage}{85 mm}
\centering
\includegraphics[height=8 cm, angle=270]{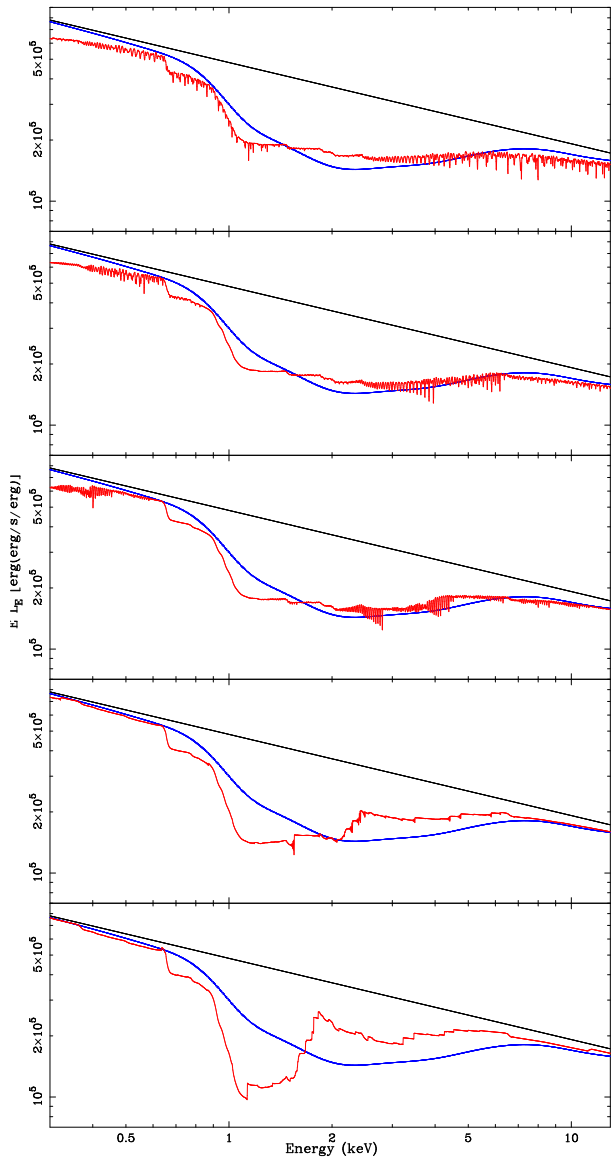}
\caption{{\small The 0.3-13 keV X-ray spectra ($E\,L_{E}$) that result from \xsc ~runs with different terminal wind velocities (red): Top to bottom: v = 0.5c, 0.7c, 0.9c, 0.95c \& 0.99c. The smeared absorption model, based on the `average' AGN properties from Middleton \etal 2007 is shown in blue and the input power-law continuum (black).}\vspace{-4mm}}
\label{vrange}
\end{minipage}
\end{figure}

The remaining parameters are set to be the same as the standard simulation parameters used in SD07. The standard set of parameters used here is thus: a 300 step chain, Log$_{10}$($\xi$)=3, N$_{H,tot}$=6$\times$10$^{22}$ cm$^{-2}$, an initial gas density of 10$^{12}$ cm$^{-3}$ and linear acceleration throughout the wind (for a more detailed discussion on the impact of other parameters, or for more information on the limitations of the \xsc ~code, see SD07).

Figure~\ref{vrange} shows \xsc ~spectra calculated with a range of terminal velocities from 0.5-0.99c. Looking past the numerical artefacts discussed above, Figure~\ref{vrange} clearly shows that a sufficiently large velocity dispersion can remove many of the sharp atomic features evident in previous \xsc ~models (SD07) and that at high terminal velocities, the resultant spectral shape is similar to that of the simplistic models presented, for example, in GD04 and SD06. Figure~\ref{vrange} clearly reveals, however, that reproducing this shape requires a terminal velocity that is considerably higher than the typical 0.3c terminal outflow velocity expected from a radiatively driven wind (even incorporating a full relativistic treatment of the wind outflow velocities - Dorodnitsyn \& Novikov 2005). Instead, the required terminal velocity is $\ge$0.9c. The only remaining plausible physical origin for a wind with such extreme relativistic velocities is material entrained by a magnetically driven jet.

\section{Conclusions}
\label{3}

GD04, and subsequent work, all clearly demonstrate that extremely large velocity dispersions are needed if the soft excess is to be explained by absorption and emission in an accretion disk wind. Yet SD07 showed that typical radiatively driven accretion disk winds do not have sufficient velocity dispersion to reproduce the smooth upturn of the soft X-ray excess at $\sim$1 keV.

In this letter we present the best available spectral models of AGN seen through a highly relativistic, accelerating, accretion disk wind, generated with \xsc. These model spectra can reproduce the shape of the smooth soft X-ray excess, however the outflow {\it must} accelerate to highly relativistic terminal velocities, of the order of $\sim$0.9c, in order to produce sufficient velocity dispersion to smear out characteristic sharp atomic features. This effectively rules out radiative driving (both line and continuum) or thermal driving of the outflow, both of which are unable to produce such high terminal velocities.

Magnetic driving is thus the only remaining, plausible, acceleration mechanism for such an outflow. The potential of magnetic fields to drive a wind is difficult to quantify, because the properties of such a wind will depend strongly on the field configuration (Blandford \& Payne 1982; Proga 2000, 2003). Some progress is being made, however, towards understanding the role of magnetic fields in accretion disks (\eg Hawley \& Krolik 2006, Kato 2007). Magneto-hydrodynamic (MHD) simulations of the self-consistent magnetic field structure of accretion disks naturally produce jets and associated outflows. The current simulations show two distinct structures; a Poynting flux dominated jet along the spin axis of the central black hole and a larger solid angle, matter dominated, centrifugally driven, funnel wall jet. The latter appears to be the most obvious candidate for the origin of a reasonable quantity of highly relativistically outflowing material. An outflow aligned approximately along the spin axis of the black hole is also attractive from a {\it geometric} perspective because soft X-ray excesses are primarily observed in type-1 AGN, which (in the current unified AGN picture) are preferentially viewed at low inclinations (see also Sims 2005), while radiatively driven accretion disc winds are often concentrated towards the equatorial plane and are thus more likely to be observed along the l.o.s in type-2 AGN.

However, the MHD simulations do not support this identification in detail, primarily because the maximum outflow velocity typical of the simulated funnel wall jets is small, $\sim$0.4c, and the jets have very low densities, $n_{jet}$$\sim$500 cm$^{-3}$ (Hawley \& Krolik 2006). Thus, not only is the velocity dispersion in such an outflow too small to produce the smearing required for the soft X-ray excess, but such low density material is also likely to be completely ionized, and have too low a column, to result in considerable amounts of absorption of the X-ray spectrum. While the numerical MHD flows are still in their infancy, and thus these numbers are not necessarily robust, it appears more likely that the soft excess does not arise from a continuous relativistic outflow as modelled here.

However, absorption may still be the origin of this spectral feature, but with clumpy material in more chaotic velocity field (Chevallier \etal 2006, SD06), especially if this material only partially covers the source, so that the equivalent width of the characteristic atomic features are further diluted by the continuum (Miller \etal 2007). Nonetheless, with reflection and continuous absorption models all apparently ruled out as mechanisms for producing the soft X-ray excess, this important component of the X-ray spectra of many AGN remains as much a mystery today as it was when it was first discovered.

\vspace{-2mm}
\section{Acknowledgements}

The authors thank Tim Kallman for writing and supporting the \xst ~code and Kris Beckwith, Jonathan Gelbord \& Omer Blaes for fruitful discussions. This research has made extensive use of NASA's Astrophysics Data System Abstract Service.  NJS and CD acknowledge financial support through a PPARC PDRA and Senior fellowship, respectively.

\end{document}